\documentclass[aps,prx,twocolumn,amsmath,amssymb,superscriptaddress,notitlepage]{revtex4-2}
\usepackage{amsmath,amssymb,bm}
\usepackage{graphicx}
\usepackage[colorlinks=true,linkcolor=blue,citecolor=blue,urlcolor=blue,bookmarks=false]{hyperref}
\begin{document}

\title{End-state-assisted tunneling at a topological insulator STM tip}

\author{Rui Chen}
\affiliation{Department of Physics, Hubei University, Wuhan 430062, China}

\author{Jie Zhou}
\affiliation{Department of Physics, Hubei University, Wuhan 430062, China}

\author{Bin Zhou}
\affiliation{Department of Physics, Hubei University, Wuhan 430062, China}

\author{Dong-Hui Xu}\email[]{donghuixu@cqu.edu.cn}
\affiliation{Department of Physics and Chongqing Key Laboratory for Strongly Coupled Physics, Chongqing University, Chongqing 400044, China}
\affiliation{Center of Quantum Materials and Devices, Chongqing University, Chongqing 400044, China}

\begin{abstract}
Recent spin-sensitive scanning tunneling experiments with topological-insulator tips motivate a microscopic understanding of how tip geometry controls current delivery. We show that topological end states mediate resonant tunneling through the gap opened by interedge hybridization in a quantum spin Hall tip. An asymmetric termination gives oppositely spin-polarized helical paths unequal effective lengths and resonance widths. Within an independent-path description, their transmitted-current imbalance vanishes at a matched resonance and becomes finite away from it. Coupling these directional channels to a $d$-wave altermagnet yields a spin-dependent interference contrast. Hybridized end states along a stepped termination provide multiple tunneling resonances. Our results establish end-state-mediated tunneling as a transport mechanism for STM with strongly confined topological tips.
\end{abstract}
\maketitle

{\color{blue}\emph{Introduction}.}---Spin-polarized scanning tunneling microscopy (STM) extends local imaging and spec-
troscopy~\cite{Binnig1987RMP,Hansma1988Science,van2010RMP} to 
magnetic contrast~\cite{Wiesendanger1990PRL}, exchange-split
electronic states~\cite{Bode1998prl}, and atomic-scale antiferromagnetism~\cite{heinze2000real}. Spin-resolved tunneling and spin-selective Andreev reflection also test candidate Majorana zero modes in superconducting chains and vortices~\cite{jeon2017distinguishing,He2014prl,Sun2016prl}. In these applications, the electronic states of the tip and their coupling to the sample determine how spin information enters the measured signal.

Compensated magnets provide a demanding application because momentum-dependent spin splitting and spin-dependent transport can coexist with vanishing net magnetization~\cite{Naka19NC,Ahn19PRB,hayami2019momentum,Yuan2020Giant,mazin2021prediction,ma2021multifunctional,smejkal2022NatRevMat}, as exemplified by altermagnetism~\cite{Smejkal22PRX,ifmmode2022Emerging}.  Recent experiments using topological insulator tips have reported bias-reversible magnetic contrast~\cite{Aishwarya2022Science,Wang2026Nature,banerjee2025axionic,YangGF2026arXiv} and spin-sensitive quasiparticle interference (QPI) in KV$_2$Se$_2$O~\cite{Wang2026Nature,YangGF2026arXiv}. The probes are interpreted in terms of distinct boundary states: helical edge or hinge channels in Bi$_4$Br$_4$~\cite{Wang2026Nature} and spin–momentum-locked surface states in SmB$_6$~\cite{YangGF2026arXiv}. 
Interpreting such measurements requires understanding how these extended boundary states supply electrons to the local tunneling junction. For a helical tip, a sufficiently narrow apex can gap the boundary channels and suppress their connection to the sample even while the wider tip remains conducting. Identifying the states that sustain tunneling through this confined region is therefore essential, because their spectrum and coupling determine the energy-dependent transmission through which the sample is probed.

\begin{figure}[!t]
	\centering
	\includegraphics[width=0.9\columnwidth]{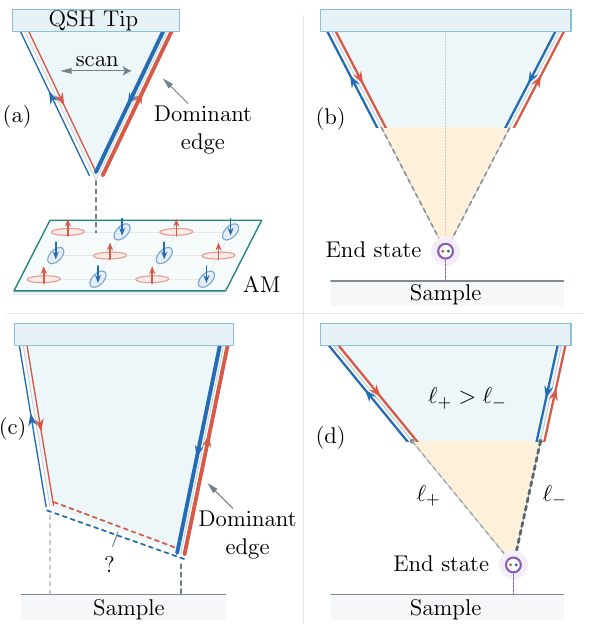}
	\caption{End-state-assisted tunneling at a confined QSH apex. (a) Helical-edge-mediated STM of
		an altermagnet (AM), motivated by the Bi$_4$Br$_4$-tip measurements~\cite{Wang2026Nature}.
		(b) Interedge hybridization opens an apex confinement gap; a localized end state (purple) mediates resonant tunneling. (c) Asymmetric termination, with the question mark denoting the microscopic channel-selection problem. 
		(d) The supplying edges couple to a common end state through unequal paths, $\ell_+>\ell_-$, producing different escape amplitudes. Yellow shading marks the hybridization-gapped region. Red and blue label ideal helical branches, with arrows specifying their propagation directions. The two supplying sides belong to the same boundary perimeter; each supports
		counterpropagating Kramers partners.
		}
		\vspace{-0.5cm}
	\label{fig_device}
\end{figure}

A quantum spin Hall (QSH) tip provides a minimal setting for addressing this problem [Fig.~\ref{fig_device}(a)]. When its terminal width becomes comparable to the edge-state penetration length~\cite{Li2011PRB,JiangHua2014}, opposite edges hybridize and open a gap~\cite{Zhou2008PRL,Shen2017TI}, suppressing direct subgap transmission. Localized states supported by the termination can instead connect the conducting edges to the sample through resonant tunneling [Fig.~\ref{fig_device}(b)]. Their coupling to the supplying channels depends on the boundary geometry. Since ideal helical modes on opposite edges carry opposite spins at a fixed propagation direction~\cite{brune2012spin}, unequal path transmissions can produce a directional spin polarization. Time-reversal symmetry nevertheless ensures that equally occupied Kramers partners give zero equilibrium spin density. Whether the directional imbalance appears in a charge measurement therefore depends on how the contact selects the helical contributions [Fig.~\ref{fig_device}(c)].

In this Letter, we show that topological end states mediate resonant tunneling through a strongly confined QSH apex. Row-end orbitals supplied by dimensional reduction~\cite{Klaassen2025NC} hybridize along a stepped termination and acquire substantial apex weight (End Matter A). These states provide transmission resonances within the gap that suppresses direct edge transport. For an asymmetric termination, the supplying edges reach the localized end state through unequal effective paths [Fig.~\ref{fig_device}(d)]. Their different resonance widths generate unequal transmitted weights away from resonance.  The resulting directional polarization requires the contact to preserve the association between spin and path. Coupling these channels to a $d$-wave altermagnet yields a spin-dependent interference contrast, with the shared end-state propagator connecting the path weights to the sample response. The theory thus relates the terminal spectrum and path-dependent resonance widths to STM current delivery and the spatial response of a spin-split sample.

\begin{figure}[t!]
	\centering
	\includegraphics[width=\columnwidth]{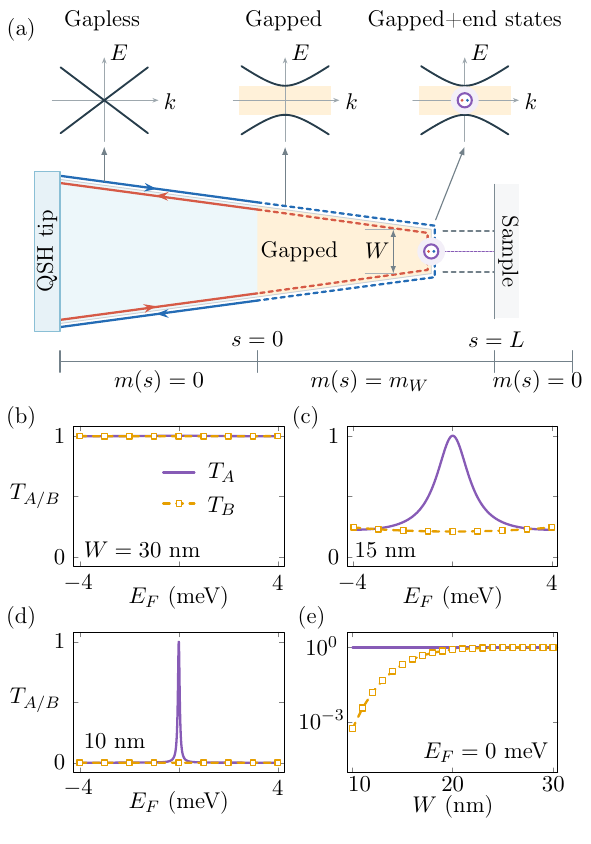}
\caption{ 
	(a) Schematic of a tapered QSH tip coupled to a sample.
	Hybridization between opposite edges opens a finite-size gap near
	the apex (yellow region), while a localized end state (purple)
	can mediate resonant tunneling through this region.
	The upper sketches illustrate the gapless edge spectrum,
	the hybridization-induced gap, and an end state inside the gap.
	Red and blue channels denote opposite spins, with arrows
	indicating their propagation directions.
	The lower profile shows the effective mass barrier:
	$m(s)=m_*e^{-W/\lambda_\psi}$ for $0<s<L$ and $m(s)=0$ outside,
	where $s$ is the longitudinal coordinate, $L=60$ nm is the barrier
	length, and $W$ is the effective transverse width of the tip.
	(b--d) Transmission versus Fermi energy $E_F$ for
	$W=30$, $15$, and $10$ nm, respectively.
	Violet solid curves show the end-state-assisted transmission
	$T_B$, while orange dashed curves with open square markers
	show the end-state-free transmission $T_A$.
	(e) Transmission at $E_F=0$ versus $W$. As the tip narrows, direct transmission is
	strongly suppressed, whereas the end-state-assisted channel
	retains unit transmission at resonance.}
	\label{fig_finite_size_effect}
\end{figure}

{\color{blue}\emph{Model}.}---We employ a minimal four-band QSH model to capture the helical edge transport and its modification at the atomic-scale apex~\cite{Shen2017TI,Kane2005PRL2,Bernevig06PRL,Konig07sci,Qi2006PRB,LuHZ2010PRB,Lu13prl-QAH}
\begin{align}
	H_{A/B}(k)=&\left[M_{A/B}-\frac{4B}{a^2}+\frac{2B}{a^2}(\cos k_xa+\cos k_ya)\right]\tau_z\sigma_z \nonumber \\ 
	&-\frac{\gamma}{a}(\sigma_y\sin k_xa-\sigma_x\sin k_ya), 
\end{align}
where $k^2 = k_x^2 + k_y^2$ and $k_\pm = k_x \pm i k_y$. $\sigma_i$ and $\tau_i$ are Pauli matrices acting
in the spin and orbital spaces, respectively. In the numerical calculations, we take $B=-30$ eV$\cdot$\AA$^2$, $\gamma=3$ eV$\cdot$\AA, and the lattice constant $a=1$ nm~\cite{Lu13prl-QAH,LuHZ2010PRB,Chang2013Science}.

To isolate the contribution of terminal states, we compare two QSH parameter regimes with matched confinement gaps.
We take $M_A=-0.05\,\mathrm{eV}$, associated with the widely studied (Bi,Sb)$_2$Te$_3$ model~\cite{Lu13prl-QAH,LuHZ2010PRB,Chang2013Science}, as an end-state-free regime, and $M_B=-1.15\,\mathrm{eV}$ as the end-state-assisted regime. The two models have identical transverse probability profiles and finite-size gaps at $k_x=0$, with the same leading linear edge dispersion;
their finite-$k_x$ corrections differ (Secs.~SI and~SII of the
Supplemental Material~\cite{Supp}). Their distinction emerges upon dimensional reduction: for the row
Hamiltonian and termination considered here, $H_A$ is trivial, whereas
$H_B$ supports localized end states protected by its chiral structure.
This distinction requires more than time-reversal symmetry alone.  For a tapered STM tip~[Fig.~\ref{fig_finite_size_effect}(a)], finite-size hybridization suppresses
the extended-edge contribution near the apex in $H_A$.
In contrast, $H_B$ supports states at the apex and step terminations
(End Matter~A), providing additional channels for resonant tunneling.

{\color{blue}\emph{Dirac quantum dot model}.}---Using the edge theory~(Sec.~SII of the Supplemental Material~\cite{Supp}), we derive an effective model for transport along one boundary of a tapered QSH tip, accounting for hybridization with the opposite boundary and coupling to a localized end state , as follows:
\begin{align}
H_A&=\int ds\,\psi^\dagger(s)
\left[-i\gamma\rho_x\partial_s+m(s)\rho_z\right]\psi(s),
\\
H_B&=H_A+\epsilon_d d^\dagger d+
V_0\psi^\dagger(s_0) d+V_0^\dagger d^\dagger \psi(s_0).
\label{eq:dirac_dot_H}
\end{align}
where $s$ is transport direction~[Fig.~\ref{fig_finite_size_effect}(a)] and $\rho_i$ act on the opposite boundaries of the QSH tip. For $m(s)=0$, the effective Hamiltonian describes a gapless Dirac spectrum, corresponding to conducting edge channels without interedge hybridization. A nonzero mass describes the gapped Dirac spectrum produced by finite-size hybridization between the opposite boundaries. 

To model the finite-size-gapped region of a tapered tip, we approximate it as a segment of uniform transverse width $W$ and longitudinal length $L$. The mass is therefore taken to be constant within $0<s<L$ and zero outside: $m(s)=m_W\Theta(s)\Theta(L-s)$, with $m_W=m_*e^{-W/\lambda_\psi}$ denoting
the width-dependent effective mass induced by hybridization
between opposite edges. We take $V_0=30~\mathrm{meV\,nm^{1/2}}$ for the end-state-assisted case, and $V_0=0$ for the 
end-state-free case. The massless regions on either side serve as conducting leads.  Here, $m_*=\frac{2|\gamma M_A|}{\sqrt{\gamma^2-4BM_A}}$ is the value of the finite gap inside the narrow segment and $\lambda_\psi=\frac{2|B|}{|\gamma|-\sqrt{\gamma^2-4BM_A}}$ is the transverse localization length of the edge-state wave functions. The width $W$ denotes the separation between the transverse hard walls. For the parameters given above, we have $\lambda_\psi=4.73~\mathrm{nm}$. We note that finite-size effects become significant when $W\sim\lambda_\psi$ and grow exponentially as $W$ decreases.

In Sec.~SIII A of the Supplemental Material~\cite{Supp}, we derive the transmission coefficients of the two systems. For the end-state-free case $H_A$ and the end-state-assisted case $H_B$, respectively, the subgap transmissions are
\begin{align}
	T_A(E_F)
	&=
	\left[
	1+
	\frac{m_W^2}{m_W^2-E_F^2}
	\sinh^2(\kappa L)
	\right]^{-1},\nonumber
	\\
	T_B(E_F)
	&=
	\left[
	1+
	\frac{m_W^2}{m_W^2-E_F^2}
	\sinh^2(\kappa L)\,
	\mathcal R(E_F)
	\right]^{-1}.
	\label{eq:transmission_AB}
\end{align}
Here, $\kappa(E_F)=\sqrt{m_W^2-E_F^2}/(\hbar v)$ is the inverse decay length inside the mass barrier, with $\hbar v=|\gamma|$. The resonant-contact factor is $\mathcal R(E_F)=[\delta+\Gamma_c E_F\tanh(\kappa L/2)/(\hbar v\kappa)]^2/(\delta^2+\Gamma_c^2)$, where $\delta=E_F-\epsilon_d$ and $\Gamma_c=V_0^2/(2\hbar v)$. When $V_0=0$, the end state decouples, giving $\mathcal R(E_F)= 1$ and $T_B=T_A$. 

It is noticed that the model possesses chiral symmetry, which pins the isolated
topological end states to zero energy in the absence of
hybridization. However, we emphasize that our tunneling mechanism does not
require zero-energy pinning and also applies to end states
at nonzero energies within the gap.

Figures~\ref{fig_finite_size_effect}(b--d) compare the transmission
with and without end-state assistance as the strip narrows from
$W=30$ to $10$ nm. Stronger hybridization between opposite edges
progressively suppresses the end-state-free transmission $T_A$.
In contrast, $T_B$ retains a pronounced resonance at zero energy,
whose peak height remains unity while its width decreases.
Figure~\ref{fig_finite_size_effect}(e) highlights this contrast:
the zero-energy transmission rapidly decreases as the strip
narrows in the end-state-free case, but remains unity in the
end-state-assisted case. These results show that a localized
end state can sustain resonant tunneling through the
finite-size-gapped region even when direct edge transmission
is strongly suppressed.

In Sec.~\ref{Sec:transport} of the End Matter, we perform numerical
transport calculations within the tight-binding approximation,
using the Landauer--B\"uttiker formula~\cite{Landauer1970Philosophical,
	Buttiker1988PRB,Fisher1981PRB} and the recursive Green's function
method~\cite{Mackinnon1985Zeitschrift,Metalidis2005PRB}.
The numerical results are qualitatively consistent with the
analytical predictions, showing suppressed transport in $H_A$
and pronounced in-gap resonant peaks in $H_B$.

\begin{figure}[t!]
\centering
\includegraphics[width=\columnwidth]{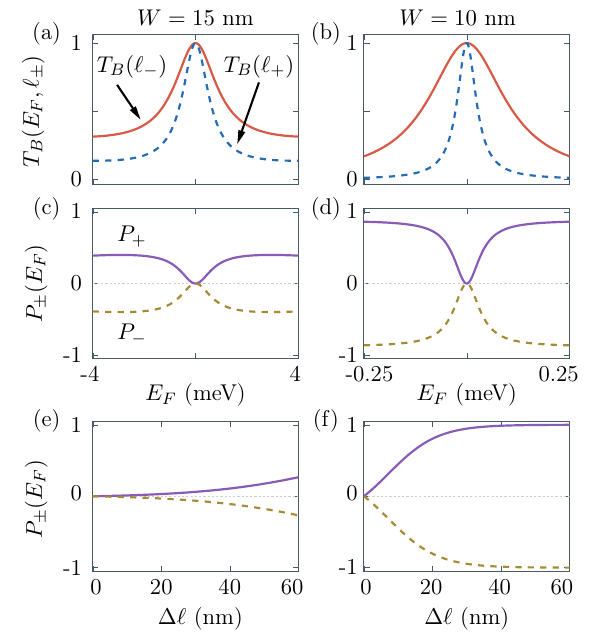}
\caption{(a,b) Transmission along the two paths for $W=15$ and $10$ nm,
	respectively, with $\ell_+=70$ nm and $\ell_-=50$ nm.
	Red solid and blue dashed curves denote $T_B(E_F,\ell_-)$
	and $T_B(E_F,\ell_+)$, respectively.
	(c,d) Corresponding spin polarizations for electron tunneling
	from the tip to the sample ($P_+$, violet solid) and in the
	reverse direction ($P_-$, ochre dashed).
	Both polarizations vanish at the zero-energy resonance.
	The two columns in (a--d) use different energy scales.
	(e,f) Polarizations versus the path-length difference
	$\Delta\ell=\ell_+-\ell_-$ for $W=15$ and $10$ nm, respectively,
	evaluated away from resonance.
	The shorter path is fixed at $\ell_-=\ell_0=50$ nm,
	while $\ell_+=\ell_0+\Delta\ell$ varies. We take $E_F=0.2$ meV in (e) and (f).}
\label{fig:asymmetric_spin}
\end{figure}

\begin{figure*}[t!]
	\centering
	\includegraphics[width=\textwidth]{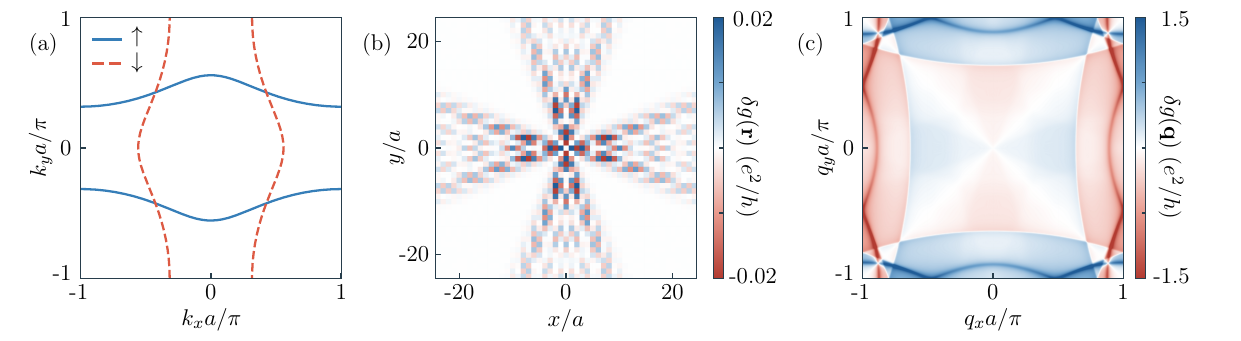}
	\caption{Spin-dependent conductance difference of an altermagnet
		probed by an asymmetric QSH tip.
		(a) Spin-up (red solid) and spin-down (blue dashed)
		constant-energy contours at $E_F=0.12t$.
		(b) $\Delta g(\mathbf{r})=g_{V_+}(\mathbf r)-g_{V_-}(\mathbf r)$
		on a $49\times49$ lattice window centered on a scalar impurity.
		(c) Real part of the unnormalized Fourier transform
		$\Delta\widetilde g(\mathbf q)$. 
		Sample parameters are $W=10$ nm, $t=100$ meV, $t'/t=-0.25$, $\mu/t=-1$,
		$\Delta_d/t=0.8$, $U/t=2$, and $t_c/t=0.06$; the
		tip-relative energy is $\omega=0.12$ meV.
		The peak $|\Delta g|$ is $0.127\,e^2/h$.}
	\label{fig:qpi_spin_tip}
\end{figure*}

{\color{blue}\emph{Asymmetric boundaries and directional spin
		polarization}.}--- The Bi$_4$Br$_4$ experiment associates spin selectivity
	with tunneling dominated by one helical edge~\cite{Wang2026Nature}. In
	the present model, geometric asymmetry gives the two
	edge-to-end-state paths unequal effective lengths. The resulting resonance widths differ, so that the shorter
	path contributes more strongly away from resonance.
	End-state-assisted tunneling thus connects the asymmetry of the supplying paths to their transmitted-current
	imbalance.

To quantify this mechanism, we describe tunneling from the two
helical edges in Fig.~\ref{fig_device}(d) using effective paths
of lengths $\ell_+$ and $\ell_-$, with mass profiles
$m_\pm(s)=m_W\Theta(s)\Theta(\ell_\pm-s)$. Within an independent-path approximation, the transmission
coefficients along the two edge-state paths are
$T_B(E,L=\ell_+)$ and $T_B(E,L=\ell_-)$, respectively,
as given by Eq.~\eqref{eq:transmission_AB}.

For propagation from the tip toward the sample, the incoming branches on the $\ell_+$ and $\ell_-$ paths carry opposite spin projections, $\downarrow$ and $\uparrow$  Fig.~\ref{fig_device}(d).
For equal incident fluxes and a contact that preserves these directional contributions, the independent-path approximation gives the energy-resolved polarization
\begin{equation}
	P_+(E)=
	\frac{T_B(E,\ell_-)-T_B(E,\ell_+)}
	{T_B(E,\ell_+)+T_B(E,\ell_-)}.
	\label{eq:positive_bias_polarization}
\end{equation}
For reciprocal paths at the same energy and scattering potential, reversing the selected direction exchanges the spin labels, giving $P_-(E)=-P_+(E)$.

Figures~\ref{fig:asymmetric_spin}(a)-\ref{fig:asymmetric_spin}(b) compare the transmissions for $\ell_+=70\,\mathrm{nm}$
and $\ell_-=50\,\mathrm{nm}$ at $W=15$ and $10\,\mathrm{nm}$.
Both paths reach unit transmission at the matched zero-energy resonance.
Away from resonance, the shorter path retains a larger transmission
because its resonance is broader.
This difference produces the spin polarizations in Figs.~\ref{fig:asymmetric_spin}(c)-\ref{fig:asymmetric_spin}(d).
The shorter path carries spin-up electrons toward the sample,
giving $P_+>0$; the reciprocal selected current has the opposite
polarization.
At resonance, equal path transmissions give $P_\pm=0$.
Reducing $W$ narrows both peaks and the central
region of weak polarization, while increasing the polarization
magnitude away from resonance.

To examine the effect of geometric asymmetry, we fix the
shorter path at $\ell_-=\ell_0=50$ nm and increase the longer
path as $\ell_+=\ell_0+\Delta\ell$, with $\Delta\ell\geq0$.
Figures~\ref{fig:asymmetric_spin}(e,f) show the resulting
polarizations for $W=15$ and $10$ nm, respectively.
Equal path lengths give equal transmission weights and
zero polarization. As the length difference increases,
the longer-path contribution is progressively suppressed,
and the polarization magnitude grows.
Over the plotted range, this increase is modest for $W=15$ nm,
whereas the $W=10$ nm case approaches $P_+=1$ and $P_-=-1$.
These results illustrate how unequal path lengths can produce
single-edge-dominated tunneling through the localized end state.

{\color{blue}\emph{Spin-polarized quasiparticle interference in an altermagnet}.}---We simulate the scanning conductance of a QSH tip
coupled to a collinear $d$-wave altermagnet, with the spin axis
parallel to the sample N\'eel vector. The sample Hamiltonian is
\begin{align}
	H_{\rm AM}
	=\sum_{\mathbf k}c_{\mathbf k}^\dagger
	\left[\varepsilon_{\mathbf k}s_0+d_{\mathbf k}s_z\right]
	c_{\mathbf k}+U\sum_\sigma
	c_{\mathbf 0\sigma}^\dagger c_{\mathbf 0\sigma},
	\label{eq:qpi_sample_hamiltonian}
\end{align}
where $c_{\mathbf r\sigma}$ annihilates an electron on the square
lattice of spacing $a$, $s_i$ act on physical spin,
$\varepsilon_{\mathbf k}=-2t(\cos k_xa+\cos k_ya)
-4t'\cos k_xa\cos k_ya-\mu$, and
$d_{\mathbf k}=\Delta_d(\cos k_xa-\cos k_ya)$, which is chosen to mimic the $d$-wave altermagnetic band
structure of the experimentally studied material KV$_2$Se$_2$O~\cite{Wang2026Nature}.

Figure~\ref{fig:qpi_spin_tip}(a) displays the characteristic
$d$-wave altermagnetic spin texture.
The spin-up and spin-down constant-energy contours extend
along the $k_x$ and $k_y$ directions, respectively, and are
related by a $90^\circ$ rotation.
The spin splitting therefore changes sign under this rotation
and vanishes along the diagonals $k_x=\pm k_y$, where the
two contours intersect.
This momentum-dependent, alternating spin polarization
coexists with zero net spin polarization after momentum
integration, illustrating a key feature of altermagnetism.
It also produces distinct spin-resolved scattering patterns
that can be probed by the directionally polarized tip.

We describe the tip using four effective Dirac path models,
indexed by spin $\sigma=\pm1$ and boundary $b=\pm$:
\begin{align}
	h_{\sigma b}
	=-i\sigma\hbar v\rho_x\partial_s+m_b(s)\rho_z,
	\label{eq:qpi_four_path_model}
\end{align}
with $m_b(s)=m_W\Theta(s)\Theta(\ell_b-s)$.
Here $\sigma=+1$ ($-1$) denotes spin up (down), while
$b=+$ ($-$) labels the upper (lower) boundary path.
The two boundaries are distinguished by their mass-segment
lengths $\ell_+$ and $\ell_-$. We retain the localized end state
and its coupling to the paths as in
Eq.~\eqref{eq:dirac_dot_H}. At scanning position $\mathbf r$,
the end state couples to the sample through $
	H_{\rm tun}(\mathbf r)
	=\sum_\sigma
	\left(t_c d_\sigma^\dagger c_{\mathbf r\sigma}
	+\mathrm{H.c.}\right)$, 
where $d_\sigma$ annihilates the spin-$\sigma$ end-state
component shared by the two paths. 
We incorporate the continuous paths and the impurity-containing
lattice sample through their self-energies on the end state.
The channel transmissions $\mathcal T_{\sigma,b}(\mathbf r)$
are calculated from the resulting Green function, retaining
the contact-induced energy shifts and broadenings. The
self-energy construction and numerical details are given in
Sec.~SIII B of the Supplemental Material~\cite{Supp}.

To connect with Eq.~\ref{eq:positive_bias_polarization}, we compare two directional channel sets at a
common energy $E$. The forward set contains the spin-up lower-boundary channel $(\uparrow,-)$ and the spin-down upper-boundary channel $(\downarrow,+)$, associated with the right-pointing arrows in Fig.~\ref{fig_finite_size_effect}(a). The reciprocal set contains $(\downarrow,-)$ and $(\uparrow,+)$.
All return channels remain in the end-state self-energy.
For each prescribed selection, the channel-resolved conductance
coefficient is
\begin{align}
	g_{V_+}(\mathbf r)
	&\equiv\frac{e^2}{h}
	\left[\mathcal T_{\uparrow,-}(\mathbf r)
	+\mathcal T_{\downarrow,+}(\mathbf r)\right],
	\notag\\
	g_{V_-}(\mathbf r)
	&\equiv\frac{e^2}{h}
	\left[\mathcal T_{\downarrow,-}(\mathbf r)
	+\mathcal T_{\uparrow,+}(\mathbf r)\right].
	\label{eq:qpi_directional_conductance}
\end{align}
The relation to complete reservoir injection is specified in
Sec.~SIII~B of the Supplemental Material~\cite{Supp}.

Scanning yields
$\Delta g(\mathbf{r})=g_{V_+}(\mathbf r)-g_{V_-}(\mathbf r)$
and its Fourier transform
$\Delta\widetilde{g}(\mathbf{q})
=\sum_{\mathbf{r}}e^{-i\mathbf{q}\cdot\mathbf{r}}\Delta g(\mathbf{r})$.
Figure~\ref{fig:qpi_spin_tip}(b) shows alternating positive
and negative conductance modulations around the impurity.
The real-space contrast and signed Fourier features
in Fig.~\ref{fig:qpi_spin_tip}(c) reverse sign under a $90^\circ$ rotation, consistent with the exchange
of the spin-up and spin-down dispersions in
Fig.~\ref{fig:qpi_spin_tip}(a).

The contrast combines unequal helical-channel weights with the
spin-dependent sample response. Comparison with opposite-bias STM maps additionally requires the bias dependence of occupations and level alignment.

{\color{blue}\emph{Role of topological end states}.}---An isolated trivial localized state can also mediate resonant
tunneling, but its contribution is concentrated near a particular
resonance whose energy depends on the local confining potential.
The advantage of topological end states in the present geometry
lies in their repeated occurrence at the appropriate boundary
terminations. For the symmetries and terminations considered here,
the nontrivial one-dimensional structure obtained by dimensional
reduction of $H_B$ supports end-state orbitals at the row ends.
The stepped tip boundary therefore naturally provides multiple
such orbitals without requiring a separate local defect at
each termination.

\begin{figure*}[h!]
	\centering
	\includegraphics[width=\textwidth]{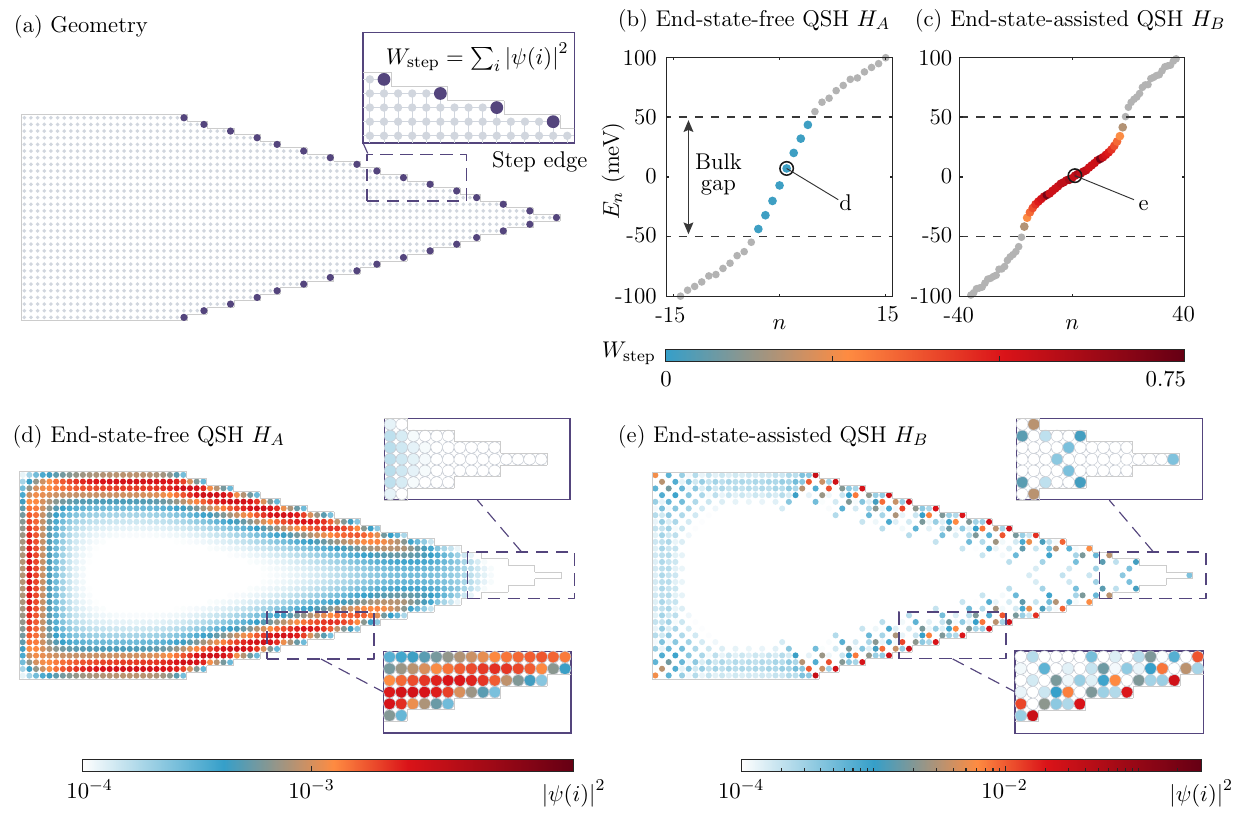}
	\caption{Spatial and spectral signatures of step-end states
		in a tapered QSH tip.
		(a) Geometry used for both $H_A$ and $H_B$.
		Purple sites mark the row terminations along the stepped
		boundary, including the apex, and define the set $\mathcal S$.
		The inset enlarges a section of the stepped boundary.
		(b,c) Energy spectra of $H_A$ and $H_B$, respectively.
		Dashed lines mark the bulk band edges at $\pm50$ meV.
		In-gap states are colored by their total probability
		$W_{\rm step}$ on $\mathcal S$, while states outside the
		gap are shown in gray.
		The eigenvalues are ordered by energy, with $n=1$ denoting
		the lowest positive-energy state.
		(d,e) Spatial probability distributions of the circled states
		in (b,c), respectively.
		The insets enlarge the apex and a section of the stepped
		boundary. Colors encode the local probability $|\psi(i)|^2$,
		with separate color scales for (d) and (e);
		each inset uses the scale of its parent panel.}
	\label{fig:endmatter_step_states}
\end{figure*}

These end-state orbitals hybridize with one another and mix
with extended edge states, producing a set of in-gap levels
with substantial step-end weight distributed over a broad
energy interval, as shown in
Fig.~\ref{fig:endmatter_step_states} of the End Matter.
When coupled to the conducting tip channels and the sample,
the hybridized states with appreciable contact overlap can
each contribute a tunneling resonance.
The resulting collection of resonances provides multiple
opportunities for transmission through the finite-size-gapped
region as the energy alignment varies.
This mechanism offers a route to tunneling that is less
dependent on the precise alignment of a single localized
level, providing a practical advantage of topological
end states in a stepped QSH tip.

\begin{acknowledgments}
	We would like to thank Zhenyu Wang and Chui-Zhen Chen for helpful discussions. This work was supported by the NSFC (under Grants Nos.~12474151, 92565103, 12304195, 12674198, 12547101, and U25D8012). D.H.X. acknowledges the Beijing National Laboratory for Condensed Matter Physics (No. 2024BNLCMPKF025) and the Fundamental Research Funds for the Central Universities (Grant No. 2025CDJIAISYB-032). R.C. acknowledges the Chutian Scholars Program in Hubei Province, the Hubei Provincial Natural Science Foundation (Grant No. 2025AFA081), the key project of Hubei provincial department of education (under Grant No. D20241004), and the original seed program of Hubei University. B.Z. acknowledges the Wuhan city key R\&D program (under Grant No. 2025050602030069).
\end{acknowledgments}

\bibliographystyle{apsrev4-1-etal-title_6authors}
\bibliography{refs-transport}

\vspace{2cm}
\onecolumngrid

\twocolumngrid
\section*{End Matter}
\subsection{Step-end states in a tapered tip}
\label{Sec:wave_Tip}
To examine the spatial origin of end-state-assisted tunneling,
we compare the spectra and wave-function distributions of
$H_A$ and $H_B$ using the same tapered geometry shown in
Fig.~\ref{fig:endmatter_step_states}(a).
The purple sites mark the terminal sites of the lattice rows
along the stepped boundary, including the apex.
We denote this set by $\mathcal S$ and quantify the probability
concentrated on these sites as
\begin{equation}
	W_{\rm step}=\sum_{i\in\mathcal S}|\psi(i)|^2,
\end{equation}
where $\psi$ is a normalized eigenstate and $|\psi(i)|^2$
is the probability on site $i$, summed over the internal
degrees of freedom.

Figures~\ref{fig:endmatter_step_states}(b) and \ref{fig:endmatter_step_states}(c) reveal a clear
contrast between the two models.
Within the bulk gap, the edge-derived states of $H_A$ carry
little weight on the step-end sites.
In contrast, $H_B$ exhibits more in-gap states with substantial
probability concentrated at the step ends.
This distinction is directly visible in the lowest
positive-energy states shown in
Figs.~\ref{fig:endmatter_step_states}(d) and \ref{fig:endmatter_step_states}(e).
For $H_A$, the probability is distributed along the boundary,
with little weight at the row terminations or the apex.
For $H_B$, pronounced local concentrations appear at the
step ends, including finite weight at the apex.

The $H_B$ levels with appreciable step-end weight span
approximately $82.9$ meV.
Their spread in energy reflects hybridization among the
row-end orbitals and their mixing with extended edge states,
as analyzed in the Supplemental Material~\cite{Supp}.
The stepped boundary thus supports a set of end-state-derived
levels distributed across a broad portion of the bulk gap.
These states can provide resonant tunneling channels when
a step termination or the apex is coupled to the sample.


\subsection{Transport}
\label{Sec:transport}
\begin{figure}[t!]
	\centering
	\includegraphics[width=\columnwidth]{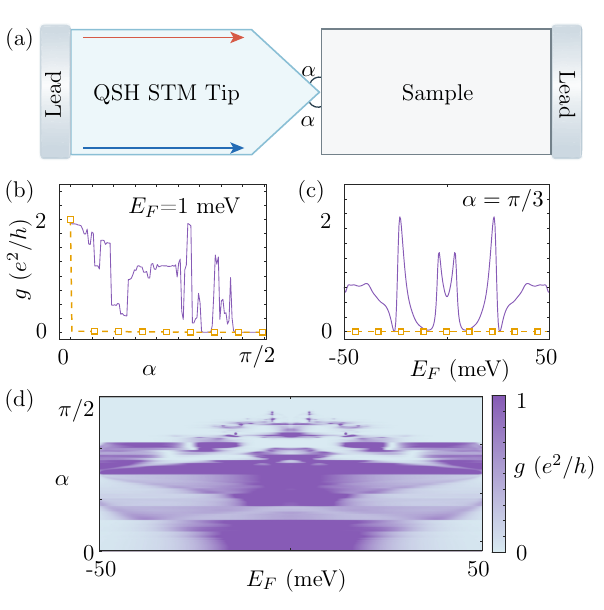}
	\caption{(a) Geometry of the QSH STM tip and sample used in the tight binding calculations. (b) Two-terminal conductance $g$ as a function of the 
		tip angle $\alpha$ at a fixed Fermi energy $E_F=1~\mathrm{meV}$.
		(c) Conductance $g$ as a function of the Fermi energy $E_F$ 
		at a fixed tip angle $\alpha=\pi/3$.
		(d) Conductance map in the $(E_F,\alpha)$ plane for the 
		end-state-assisted QSH model.}
	\label{fig_transport}
\end{figure}

We investigate the quantum transport properties of the system based on the
specific tip geometry shown in Fig.~\ref{fig_transport}(a). The two-terminal conductance is calculated by using
the Landauer-B\"uttiker formula~\cite{Landauer1970Philosophical,
	Buttiker1988PRB,Fisher1981PRB} combined with the recursive Green's
function method~\cite{Mackinnon1985Zeitschrift,Metalidis2005PRB}  (see Sec. SIII C of Supplemental Material~\cite{Supp} for more details). The setup
consists of a tapered QSH STM tip and a flat QSH sample. The angle
$\alpha$ controls the geometry of the tip apex.

As shown in Fig.~\ref{fig_transport}(b), when $\alpha=0$, both QSH models 
exhibit an identical quantized conductance of $g = 2 e^2/h$. This is because 
the uniform wide ribbon allows the helical edge modes to transport without experiencing the finite-size effect. However, as $\alpha$ becomes 
nonzero, the transport behaviors of the two models diverge sharply. The 
conductance of the end-state-free QSH model is immediately 
suppressed once the tip begins to taper. This behavior reflects the fact that 
the helical edge mode is interrupted by the large finite-size gap 
opened at the atomic-scale apex and therefore cannot tunnel  into 
the sample. In sharp contrast, the end-state-assisted QSH model 
retains a finite conductance over a broad range of $\alpha$, despite having 
the same wide-region edge dispersion as the end-state-free model. In Sec. SIII C of the Supplemental Material~\cite{Supp}, we provide further numerical results and demonstrate that this conclusion remains robust against variations in the mass parameter $M$ and tip width.

To further explore the transport behavior, Fig.~\ref{fig_transport}(c) shows the conductance as a function of the Fermi energy at a fixed angle $\alpha=\pi/3$. 
The conductance displays sharp resonant peaks inside the finite-size gap. 
These peaks are the hallmark of resonant tunneling. 
As we will show below, we attribute this resonant transport to the emergent end states at the tip apex. 
This feature is further supported by the full conductance map in the $(E_F,\alpha)$ plane for the end-state-assisted model shown in Fig.~\ref{fig_transport}(d). 
Notably, the resonant tunneling behavior becomes more pronounced when $\alpha$ is relatively large.

As $\alpha$ increases, the conductance spectrum transforms from a smooth transport window into sharp resonance branches that split across the energy range. We provide a detailed analysis of this behavior in Sec. SIII C of the Supplemental Material~\cite{Supp}. Specifically, an increase in $\alpha$ geometrically extends the length of the hybridized transition region at the tip apex, which increases the spatial extent of the gapped region. This reduces the lifetime broadening of the in-gap states, causing the resonant tunneling paths to turn into sharp lines at larger angles.

\begin{figure}[t!]
	\centering
	\includegraphics[width=\columnwidth]{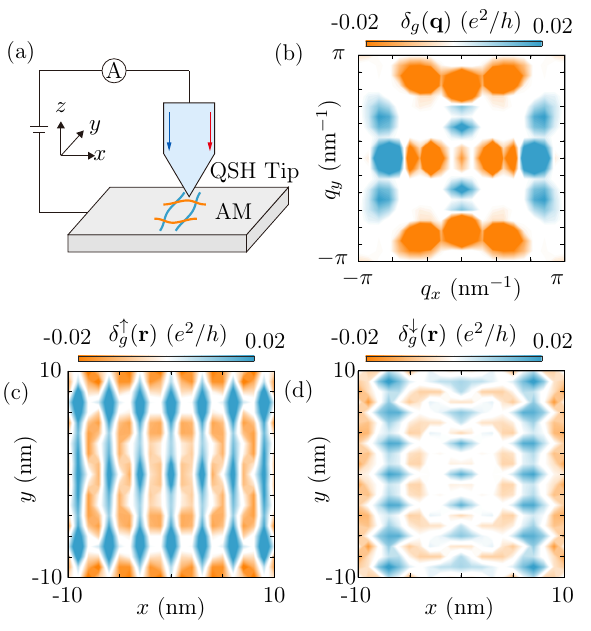}
	\caption{(a) Schematic of the qualitative scanning-transport simulation of a QSH tip coupled to a $d$-wave altermagnetic device. A QSH tip is scanned over the AM region connected to external leads.  The spin-resolved altermagnetic spectrum is shown in the device. (b) Fourier-space spin contrast  extracted from the scanning transmission maps. The opposite signs along the $q_x$ and $q_y$ directions reflect the $d$-wave altermagnetic spin texture. (c) and (d) Real-space spin-resolved transmission modulations
		$\delta_g^{\uparrow}$ and $\delta_g^{\downarrow}$, defined as the deviation
		of the calculated local transmission from its spatial average,
		$\delta_g^{\sigma}(\mathbf{r})=g^{\sigma}(\mathbf{r})-\langle
		g^{\sigma}\rangle_{\mathbf{r}}$ with $\sigma=\uparrow,\downarrow$. The two spin channels show mutually orthogonal quasi-one-dimensional modulations.}
	\label{fig_AM}
	\vspace{0.1cm}
\end{figure}

\subsection{Transport simulation}
To make a qualitative connection with the spin-resolved scanning-tunneling experiment~\cite{Wang2026Nature}, we also modeled an STM tip within the tight-binding framework and performed scanning transport calculations using the device geometry shown in Fig.~\ref{fig_AM}(a). The setup consists of an end-state-assisted QSH tip coupled to a two-dimensional $d$-wave altermagnetic  device. The microscopic Hamiltonian, the lead configuration, and the numerical implementation are given in Sec. III D of the Supplemental Material~\cite{Supp}. This geometry allows us to calculate the
two-terminal spin-resolved conductance, denoted by $g^{\sigma}(\mathbf{r})$,
as a function of the tip position $\mathbf{r}$, with
$\sigma=\uparrow,\downarrow$ denoting spin.

Figures~\ref{fig_AM}(c) and \ref{fig_AM}(d) show the spin-resolved
transport modulations, denoted by $\delta_g^{\uparrow}$ and
$\delta_g^{\downarrow}$, respectively. Here the modulation is defined by
subtracting the spatial average from the calculated transport map, $
\delta_g^{\sigma}(\mathbf{r})=g^{\sigma}(\mathbf{r})-
\langle g^{\sigma}(\mathbf{r})\rangle_{\mathbf{r}}$, where $\langle g^{\sigma}(\mathbf{r})\rangle_{\mathbf{r}}$ denotes the spatial
average of $g^{\sigma}(\mathbf{r})$ over all tip positions $\mathbf{r}$ in the
scanning region.
This subtraction removes the uniform background transport and highlights the
position-dependent scanning contrast. 

The two maps exhibit approximately
orthogonal anisotropic patterns. This behavior can be directly understood from
the spin-resolved spectrum shown in Fig.~\ref{fig_AM}(a). Its
constant-energy contours contain nearly parallel segments along orthogonal
directions. Specifically, for the spin-up sector (colored by cyan), the relevant contour segments are nearly parallel to
the $k_y$ direction and are mainly separated along $k_x$. Therefore, the
dominant scattering wave vector selected by the QSH tip is oriented along
$q_x$, with $\mathbf{q}_{\uparrow}
\simeq
(\Delta k_x,0)$. 
The corresponding real-space transport modulation can be written
schematically as $
	\delta g^{\uparrow}(\mathbf{r})
	\sim
	\cos(\mathbf{q}_{\uparrow}\cdot\mathbf{r})
	=
	\cos(q_x x)$. 
Thus, $\delta g^{\uparrow}$ varies primarily along the $x$ direction and is
relatively uniform along the $y$ direction, giving rise to the stripe-like
pattern in Fig.~\ref{fig_AM}(c). 

In contrast, for the spin-down sector, the corresponding contour segments are
rotated by approximately $90^\circ$. They are nearly parallel to the $k_x$
direction and are mainly separated along $k_y$, giving a dominant scattering
wave vector $\mathbf{q}_{\downarrow}
\simeq
(0,\Delta k_y)$. The resulting real-space response is then $
	\delta g^{\downarrow}(\mathbf{r})
	\sim
	\cos(\mathbf{q}_{\downarrow}\cdot\mathbf{r})
	=
	\cos(q_y y)$. 
Therefore, $\delta g^{\downarrow}$ varies primarily along the $y$ direction
and produces the orthogonal pattern shown in Fig.~\ref{fig_AM}(d), with strong
Fourier components mainly along the $q_y$ direction.

To further expose the momentum-space information contained in the scanning signal, we Fourier transformed the real-space transport maps and plotted the spin contrast $\delta_g(\mathbf{q})$ in Fig.~\ref{fig_AM}(b). The Fourier-space contrast shows opposite signs for the $q_x$- and $q_y$-oriented scattering features, consistent with the $d$-wave form factor of the altermagnetic spin texture. Thus, our simulation captures the same qualitative fingerprint in the experiments: spin-selective, mutually orthogonal real-space modulations and a sign-changing $d$-wave contrast in momentum space~\cite{Wang2026Nature}.

We also performed additional calculations in Sec. III D of the Supplemental Material~\cite{Supp}. First, we replaced the end-state-assisted QSH tip by an end-state-free QSH tip. In this case, the characteristic spin-selective scanning pattern disappears. The reason is that, at the atomic-scale tip apex, finite-size hybridization opens a local gap for the helical edge states. Without the localized zero-dimensional end state at the apex, the helical channel is blocked and cannot efficiently mediate tunneling into the AM device. By contrast, the end-state-assisted QSH tip supports resonant tunneling through the localized apex state, so that the helical spin selectivity survives in the scanning transport signal.

Second, we replaced the AM device by a normal-metal device while keeping the QSH tip and the lead geometry unchanged. The characteristic real-space images and Fourier-space spin contrast are again absent. This control calculation shows that the QSH tip alone is insufficient to generate the observed pattern. A normal metal does not possess $d$-wave spin splitting or symmetry-related spin-polarized $X/Y$-valley quasiparticle channels; therefore, there is no underlying altermagnetic spin texture for the helical tip to select. Any residual interference generated by the finite device geometry or by the attached leads is not locked to the alternating AM spin texture and therefore cannot reproduce the orthogonal spin-resolved maps or the sign-changing Fourier contrast.

Taken together, these control calculations support the
proposed mechanism within the device geometry and parameter
regime studied here. End-state assistance sustains tunneling
through the narrow tip apex, while the altermagnetic band
structure produces the characteristic spin-dependent
spatial modulations. The resulting scanning contrast is
qualitatively consistent with the experimental observations
\cite{Wang2026Nature}. These calculations illustrate how an
end-state-assisted QSH tip can probe altermagnetic spin
textures, while the microscopic structure of the
experimental apex remains to be established.

\subsection{Emergent end state in Bi$_4$Br$_4$}
\begin{figure}[t]
	\centering
	\includegraphics[width=\columnwidth]{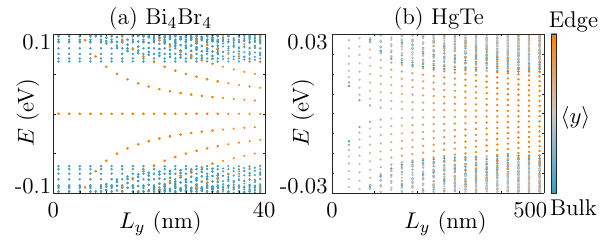}
	\caption{Energy spectra of the strip model as a function of the transverse width $L_y$. (a) and (b) represent the calculated results using the effective model of Bi$_4$Br$_4$ with $L_x=40$ nm and the realistic effective model of HgTe with $L_x=500$ nm, respectively. The green points denote bulk states, and the orange points correspond to edge states at large transverse widths $L_y$ and localized end states at small $L_y$.}
	\label{fig_BiBr}
	\vspace{0.1cm}
\end{figure}

To further verify our findings, we calculate the energy spectra using the realistic effective models for Bi$_4$Br$_4$~\cite{Zhou2015NJP,Shumiya2022NatMat} and HgTe~\cite{Konig2008JPSP,Bernevig2006Science,Konig07sci,Roth09sci}. The detailed model Hamiltonians and parameters are provided in Sec. IV of the Supplemental Material~\cite{Supp}. We observe that when the system is dimensionally reduced to a one-dimensional chain, localized end states emerge in the Bi$_4$Br$_4$ model~[Fig.~\ref{fig_BiBr}(a)]. In contrast, the HgTe model becomes fully gapped without any end states under the same one-dimensional reduction~[Fig.~\ref{fig_BiBr}(b)].

Therefore, this intrinsic property makes Bi$_4$Br$_4$ a natural candidate for realizing end-state-assisted tunneling at the extreme atomic limit, providing a possible explanation for the recent experimental observation~\cite{Wang2026Nature}. Furthermore, the dimensional reduction mechanism may provide an explanation why such topology-driven tip applications have remained elusive in prototypical topological materials like HgTe~\cite{Bernevig2006Science,Konig07sci,Roth09sci} or Bi$_2$Te$_3$~\cite{Chang2013Science,Yu2010Science}. In those  systems, the inevitable finite-size hybridization at the atomic apex trivially gaps out the edge modes without generating any intermediate end states, thereby blocking the topological transport channels.

\end{document}